# On the Potential of Social Media Data in Urban Planning: Findings from the Beer Street in Curitiba, Brazil

Santala Ville[a]*, G. Costa[b], L. Gomes-Jr[b], T. Gadda[c] and T. H. Silva[b,d*]

[a]*Geography and Sustainable Communities, University of Wollongong, Australia;*
[b]*Informatics, Universidade Tecnológica Federal do Paraná. Curitiba, Brazil;*
[c]*Civil Engineering, Universidade Tecnológica Federal do Paraná. Curitiba, Brazil;*
[d]*School of Cities, University of Toronto, Canada.*

*Corresponding authors:

Santala Ville, School of Geography and Sustainable Communities, University of Wollongong, Northfields Ave, Wollongong NSW 2522 Australia, vvs851@uowmail.edu.au;

Silva Thiago H., University of Toronto, Myhal Centre, Suite 853 55 St. George Street, Toronto, ON M5S 0C9, silva.th@utoronto.ca

# On the Potential of Social Media Data in Urban Planning: Findings from the Beer Street in Curitiba, Brazil

Social media makes available vast amounts of data for various types of analyses. Cities have the opportunity to explore this new data source to study urban dynamics and complement traditional data used for urban planning. We investigate Untappd social media data in the context of urban planning in Curitiba, Brazil. We analyze the project to create a Craft Beer Street, recently announced by the municipality to promote local beers in Curitiba, in order to study the potential of exploring social media data to support the planning of this project. Our results indicate that social media data could have helped to guide the decision of the Beer Street creation and can potentially become a strategic urban planning tool.

**Introduction**

Cities around the world have been creating initiatives to open public data. Nowadays, data sources publicly available are diverse and widespread. Despite the success of open data governmental initiatives, the amount of available data is dwarfed by the stream of data generated by citizens' interactions on social media platforms such as Instagram[1], Foursquare[2] and Untappd[3]. These examples of social media are also called location-based social networks (LBSNs), where each user, using its mobile device, acts as a sort of local sensor. Social media data might provide valuable information on the usage of space, citizens dynamics, and activities in a certain place (Cranshaw et al., 2012; Silva et al., 2014; Noulas et al., 2011; Schwartz et al., 2013; Cerrone et al., 2015).

---

[1] http://www.instagram.com.

[2] http://www.foursquare.com.

[3] http://www.untappd.com.

In that way, social media offers unprecedented data that can potentially lead to different urban planning decision-making.

Observing the city in real-time and with a more detailed spatial and temporal scale can open new opportunities to monitor and plan ongoing and future development at a neighborhood level (Cerrone et al., 2015, Silva et al., 2019). The vast amount of produced data, related to various socio-territorial aspects, may be a rich source of information supporting the decision making of individuals, businesses, and cities (Silva et al., 2016). Although social media data has attracted significant interest from governments all around the world, there are still challenges and unanswered questions related to the successful utilization of social media data in urban planning practices.

Studying urban dynamics in different spatial scales has been traditionally challenging, requiring extensive social studies, interviews, surveys, and observations, which provide a limited representation, often with a high financial cost. In this paper, we explore the potential of employing social media data to help not only mitigating these urban planning challenges but also to provide planning a new and powerful tool that can assist in the understanding of urban dynamics. Social media data can potentially help analysing how people use the urban space and provide new knowledge about urban dwellers, the urban area under the scope, and their relationships.

We explore a case study of planning a Beer Street, a designated area to consume and support the local beer brewing industry, in the city of Curitiba, Brazil. Curitiba, the largest city in the south of Brazil, is globally known for its urban planning based on the so-called transit-oriented development (TOD) as well as for being the cradle of the Bus Rapid Transit (BRT) system. In addition, Brazil is adopting digital technologies at a rapid pace, with 82% of the population with internet access and 61% of smartphone ownership (for the age group of 18-34 years[4]). Besides, in 2019 Brazil is believed to have around 140 millions of social media active users, growing around 8% from 2018 [5]. Currently, as Curitiba struggles in maintaining its status at the forefront of city planning, the city seeks for new methodologies and approaches to help the process of urban planning. We argue that social media may have an important role to play. In order to shed light into it, we explore a real world case study: the project of the so-called Craft Beer Street - a public area that the city is planning to promote local craft beers. This context is a compelling case since a previous study (Silva and Graeml, 2016) demonstrated that people in the city seem to have more interest in craft beers than in most other big cities in Brazil.

In this study, we investigate whether and how the use of social media could help urban planning under the scope of the planned Beer Street in Curitiba. Using a real case study enables us to produce empirical evidence and create understanding of real opportunities and challenges regarding the use of social media and other open data sources in urban planning. In order to do so, we firstly analyzed data from Untappd, a mobile phone application for sharing beer drinking experiences, to demonstrate how social media data could be used for identifying functional areas. Secondly, we investigated how social media data could be enriched with official open datasets, exploring citizens' complaints and requests from the city's official contact center. Additionally, we analyzed peoples' reactions in social media regarding this particular

---

[4]http://www.pewglobal.org/2016/02/22/smartphone-ownership-and-internet-usage-continues-to-climb-in-emerging-economies.

[5]https://datareportal.com/reports/digital-2019-brazil?rq=brazil

urban planning case.

Our study provides examples of ways to utilize social media data for a particular case study and how this data could assist in local decision-making. We also evaluate the challenges of working with the data collected and methodologies used. Our results indicate that in relation to the Beer Street planning case, social media data provides relevant and useful information to support decision-making and data could not only work as a planning tool up front but also provide opportunities to evaluate and legitimate decisions.

The article is organized as follows. Section 2 explores the development of evidence based planning and aspirations of using new data sources, such as social media data. Section 3 presents the case study and our empirical approach. Then Section 4 presents methodological procedures and results from the empirical case study. Results are followed by discussion, which further analyses the potential and challenges of using social media data in a specific context. Finally, the study is concluded with a few remarks and recommendations for future research.

**Data-driven Urban Planning**

Although data availability has always played an important role in planning and decision-making, recently, it has become a central requirement in the pursuit of an improved city management, alongside with the development of the 'smart city' concept. Conventional city planning requires lots of information on the anthropic as well as the bio-physical urban system. If we consider, for instance, the planning of a public transportation route, it requires very complex analysis based on, e.g., commuting patterns and demographics. Usually, gathering all the data a city needs for decision making with traditional measurements is time and labor intensive, and not rarely, also expensive. In recent years, cities and academia have started to look at new opportunities provided by developing information technology. The development of information and communication technology enables cities to collect, measure, visualize and analyze data [6] in great quantities and faster than ever.

The nature of evidence and its role in policy-making has changed considerably during different phases of planning ideologies (Davoudi, 2006). Traditionally, local authorities have paid a lot of attention to quantitative data for monitoring and planning. From a positivist point of view, information and knowledge are providing enough support for policymakers to choose the best or most optimal course of action among a range of options. In that case, the information is perceived in political processes as a value free and objective basis for making well-justified decisions (Dühr and Müller, 2012).

In a social–constructivist understanding, on the other hand, data and information are never seen as objective and unbiased inputs into planning processes (Dühr and Müller, 2012). Valid information and knowledge are always dependent on the circumstances in which each decision making takes place. In order to incorporate the increasing importance of citizens' views as well we to evaluate planning decisions, participatory methods were implemented into planning practices in the 1960s.

---

[6] In this study, data is considered the simplest form of information. Data becomes evidence as it is put into context and used to support decision making.

*Promises of Social Media Data*

Social media has changed the way we communicate with each other, and simultaneously, applications are collecting vast amounts of information on our daily life activities. As one of the main factors behind the big data movement, social media data has been gaining increasing attention and there are strong indicators that it is changing the planning of cities (Batty, 2013). Social media may increase the awareness of people, and it could also simplify the collection of public opinions and enrich the data used in the decision making process. Similarly, citizens' awareness of local policies has increased the demand for more real world and lived experience based evidence to support the decision making process (Dühr and Müller, 2012).

Notably, the rise of social media popularity provides new data and new ways to look at the city, making the value of social media more evident (Kleinhaus et al., 2015). Opportunities of social media data, especially LBSNs where each user acts as a sort of local sensor, have sparked interest in using new data sources to measure activities and the use of urban spaces (Cheng et al., 2011; Noulas et al., 2011; Cranshaw et al., 2012; Kling and Pozdnoukhov, 2012; Silva et al., 2013; Schwartz et al., 2013; Silva et al., 2014). While planning has traditionally been based on datasets limited to administrative divisions, these new data sources give cities an opportunity to implement spatial analysis at different scales. Social media can provide more sophisticated understanding of the urban territory, supporting analysis on a wider-scale, finer resolution and possibly close to 'real-time' (Kitchin, 2014).

Recently, different social media and mobile applications have been studied as tools for increasing public participation in urban planning (Ertiö, 2015; Afzalan and Evans-Cowley, 2015; Jones et al. 2015; Kleinhaus et al., 2015). The growing interest in the potential of social media and mobile technologies reflects the broad urban agenda to foster citizen participation in urban planning (Kleinhaus et al., 2015). The potential of urban data is often associated with its size and easiness to collect and analyze, but social media can also provide new opportunities for cities to hear the voice of user groups that do not formally participate in the planning and decision making process.

Social media data has been used for identifying specific activities and spaces within cities. As some studies have pointed out, e.g. Wakamiya et al. (2011) and Frias-Martinez and Frias-Martinez (2014), it is possible to use social media data to study and define the characteristics of certain urban areas. Data shared by users in systems such as Twitter or Foursquare can be used for analyzing and clustering selected groups and activities in the city, which eventually makes it possible to characterize urban areas and differentiate them (Silva et al., 2014). This kind of analysis can provide city planners with detailed information on the usage of space without being forced to analyze phenomena in the city through administrative divisions (e.g. neighborhood division).

Previous studies have shown that social media, especially LBSNs, may be used to deepen the understanding of user behavior and the city dynamics (Zhang et al., 2013; Noulas et al., 2011; Preoțiuc and Cohn, 2013; Wu et al., 2014; Cranshaw et al., 2012; Kling and Pozdnoukhov, 2012). The consensus seems to be that studying the city through social media data offers an important and easy way to understand both people's activity and interests (Schwartz et al., 2013). Social media data also brings new valuable data to study human mobility patterns (Wu et al., 2014; Cheng et al., 2011). In addition, social media data can be used to discover functional areas inside the cities (Vaca et al., 2015; Wakamiya et al., 2011).

There are some studies on the use of social media as a new source of data to understand the behavior of urban dwellers and the usage of urban space (Cerrone et al., 2015; Xia et al., 2014; Afzalan and Evans-Cowley, 2015; Kavanaugh et al., 2012). This indicates that there is a growing interest in using LBSNs data for urban planning, however, more detailed case studies are needed to fully understand the opportunities and challenges of using social media for urban planning purposes. Our study presents evidence based on a real life example, emphasizing: (i) how social media data helps in identifying functional areas, (ii) how the findings could be used by the city to improve planning practices, and (iii) how social media can help to understand the reactions of people regarding a urban planning decision.

**Case study: Planning the Beer Street in Curitiba**

Especially since the 1990's the city of Curitiba became well-known around the world for its urban planning based on TOD. Within Brazil Curitiba has been among the best cities to live. More recently, Curitiba became also nationally recognized as one of the most important cities in terms of production and consumption of craft beer, and has been branded as the Brazilian craft beer capital by local authorities[7].

In 2017, the city of Curitiba announced a plan to create a Craft Beer Street in the city. The Beer Street will be at Hauer neighborhood, more specifically at Carlos de Laert Street, where craft beer tastings take place every Friday and Saturday. The central idea of the Beer Street is to encourage the craft beer pole already installed in the city, while providing an extra tourist attraction. In order to conclude the project, the city plans to hold test events on the target street to help to understand the needs and impacts for the surrounding population[8].

---

[7] https://goo.gl/NzuF65, accessed April 5th 2020.

To investigate the actual planning decision and explore the potential new data sources might offer, we employed data from three different sources: (i) Untappd, (ii) citizen complaints from Curitiba's Open Data Portal and (iii) readers' comments on online news regarding the announcement of the beer street initiative. Exploring the new data sources and combining them to existing official datasets provides a fruitful chance to explore opportunities and limitations that these new social media data sources have regarding urban planning, and where they might fit in future planning practices.

Data produced via Untappd has been previously explored for understanding regional and social differences in consumption. For instance, Silva and Graeml (2016) collected and analyzed messages generated by Brazilian users and reported that users from some cities in the country have developed a more sophisticated taste for especially craft beer. Chorley et al. (2016) also studied data from Untappd in order to understand the drinking habits of people through the lenses of social media. According to the study, one of the advantages of using Untappd data is that a formal schema is used for all check-ins (data shared by users), which makes them easy to explore for data analysis.

In addition to Untappd data, we further analyze beer consumption areas by combining the initial results with officially produced public data that has been made openly available through Curitiba's Open Data Portal (dataset Reports). We use the so-called Reports data, which is location-based data generated on citizens' reports to the municipality of Curitiba. This dataset combined with the data received from social media can help the city to better understand the characteristics of a certain areas.

---

[8] https://www.bandab.com.br/geral/curitiba-devera-ter-em-breve-a-rua-da-cerveja-saiba-onde-ela-ficara/

Besides providing data to support administrative decisions, social media also offers the opportunity of assessing citizens' reactions in almost real time. When the press or the government itself publishes news related to city planning, a stream of comments is produced in the news and social media sites. The Beer Street project in Curitiba is a good example of the value of social media interactions in assessing citizens' concerns. Even though the project is, as of the writing of this paper, in the initial phases of planning, social network interactions generated a good amount of comments from diverse sources. The first meeting to discuss the creation of the Beer Street was reported by several news outlets and posted on social media. In a matter of hours following the news, hundreds of comments were posted by citizens expressing concerns or support regarding the initiative.

**Results**

*Identifying Popular Beer Consumption Areas*

We collected usage data from the Untappd application for two different periods: from 06 April 2013 to 30 April 2013 (Dataset 2013); and from 03 November 2016 to 05 March 2017 (Dataset 2017). We considered only one check-in per user in a specific drink location to avoid bias in the data caused by heavy users. Besides that, we also considered only messages containing the type of beer being consumed and geographic information about the location. In total, we kept 400 records for Dataset2017 and 71 records for Dataset2013.

The first step in our analysis was to discover popular areas in the city based on the number of check-ins observed in our datasets. Identifying these areas is useful because the target consumers are already visiting these places. Therefore, this approach may increase the chances of success of the new development. In this study, we propose to use the clustering algorithm DBSCAN (Ester et al., 1996). For this algorithm, we set the eps parameter as 250 meters and considered minPts=10 for all datasets and also minPts=5 for the Dataset2013. Since the check-in points are located with latitude and longitude, to calculate the distances we considered in this step the great-circle distance (using the Haversine Formula).

Figure 1 shows the results of the clustering results for Dataset2017. We can see that the algorithm found eight clusters, all of them were labeled with letters. The area where the Beer Street is planned to be built is identified (see cluster H). The similar analysis for Dataset2013, only identified cluster H, which indicates that the Beer Street area has been popular since 2013. It is worth highlighting that the area for the planned Beer Street was identified by the approach in all analysis performed in this study.

[Figure 1 around here]

If the decision makers of Curitiba were using the strategy we are discussing to help make better urban planning decisions, at this moment they would have those eight candidate areas for the creation of the Beer Street. The first observation is that all these clusters are important areas associated with free time socialising and craft beer consumption, and it emerged from the user data that implicitly express their preferences. Each of them deserving further analyses by the administration.

*Improving the Understanding About the Areas of Craft Beer Consumption*

In order to understand key characteristics of each identified area, we explore the open dataset Reports that contains various citizens' reports to the Central 156 contact center --- the main communication channel between the citizens and the municipality of Curitiba. Reports are available from June 2016 to March 2017. Each report on the dataset includes, among other attributes, an ID, a type, a subject, a subdivision, date and time when a user contacted the service, citizen's comment and address occurrence of the issue.

In possession of these data, we created a buffer of radius 200 meters around each point from the identified clusters. Buffers from the same cluster were dissolved so that, in the end, each cluster was represented by only one buffer. There is no overlap between clusters. Next, we filtered resulting reports containing subdivisions relevant to our case study and grouped them into meta-categories of problems. This was necessary because the original dataset has several subjects and subdivisions that were not very distinct from each other. The resulting meta-categories are described in Table 1. After all these steps, we were left with 3,616 reports.

[Table 1 around here]

Before we proceed, it is important to mention some of the key features that a street should have to support a beer street creation: F1) it must support being blocked on weekends for several hours without causing a considerable impact on traffic; F2) it cannot be located nearby calm zones, such as hospitals and residential areas; F3) it cannot affect negatively on current businesses on it; F4) it must be large enough to support one or more stages for concerts, products exposition, food and beer commercialization, and fit comfortably a public of at least 10k people[9]. Curitiba has a commission for the analysis of events that has 13 representatives from different sectors, including Urban Planning Secretary, Health Secretary, Social Defense Secretary, Transport Secretary, and Chamber of Commerce. This commission is responsible to concede event permits. While there are no public guidelines for the specifications of events, we believe that these requirements are among the most important ones.

Figure 2 shows the number of reports for each category in each cluster. Note that Clusters *C* and *D* are the most problematic, with the most significant number of reports per 100m$^2$. Together, they represent about 84% of the total reports. In addition, *Parking Infractions* is the most common report in all the clusters. Clusters *A*, *F* and *G* are all located in busy streets, and finding a parking place in those areas is usually difficult. This might represent an incentive for people to park in prohibited places. *A*, *F* and *G* may not be good candidates for the Beer Street due to possible contradictions on selection criteria *F1*: closing streets in these areas would considerably impact traffic.

[Figure 2 around here]

After carefully studying the areas of the clusters, Table 2 summarizes all problems found on them. Marked cells represent that a cluster presents a certain problem or does not meet a certain feature. The observations suggest that one of the best candidates to accommodate a Beer Street is cluster *H*. However, note that there are important issues observed in that area that should be carefully studied by the city.

---

[9] Based on an average number of public of some recent craft beer festivals promoted in Curitiba: https://goo.gl/a3OxwK.

[Table 2 around here]

*Citizens' Reactions Regarding the Beer Street*

This section concentrates on demonstrating how to evaluate the repercussions of a planning decision using social media. We explore the citizens' reactions to the Beer Street news expressed in comments on social platforms. A reaction is quantified by the comment sentiment polarity, i.e., if the overall text is positive or negative, evaluated by sentiment analysis (Serrano-Guerrero et al., 2015; Gonçalves et al., 2013). Automatic sentiment analysis gives decision-makers the opportunity to gauge the general perception of the population concerning the planning.

In the data acquisition step, we identified that most reactions to news about the project were registered on Facebook. Therefore, we focused on the most popular (based on the number of user reactions) public news about the Beer Street shared on the website. For the sentiment analysis, we employed SentiStrength, a publicly available tool (Thelwall et al., 2010). The mechanism can report, for each analyzed text, a sentiment polarity value ranging from -4 (strongly negative) to +4 (strongly positive). Sentiment analysis in this context is, however, especially challenging. Political bias often contaminates the comments producing reactions expressing personal views about the political figures that are not necessarily related to the topic of discussion. To handle this type of issue, we employed information retrieval (Manning et al., 2008) techniques to assign a relevance score to each comment, based on the relevance to the original news article.

Figure 3 shows the distribution of sentiment polarities for the comments studied in this work. The distribution shows a slight negative bias (mean sentiment strength is -0.05). In our application scenario, such results would prompt decision makers to investigate the comments to assess citizens' concerns. It is, however, impractical to read each of the hundreds (and, potentially, millions) of comments individually. Besides, in this scenario, there is often politically charged discourse that bears little relevance to the question.

To rank comments based on their relevance to the topic, we first collected the base text from all the news about the project that were posted on Facebook. The concatenation of the articles (eight in total) was used as a representative of our topic. We preprocessed the topic representation and each comment with traditional Natural Language Processing normalization techniques (tokenization, stemming, and stop-word removal) (Manning et al., 2008). We then used TF*IDF term weighting to identify the most important terms for the description of the topic. The important terms included the names of the participants of the initial planning meeting, name of the neighborhood considered, and terms like craft, beer, and street.

[Figure 3 around here]

Finally, we calculated the similarity between comments and the topic as a marker for relevance. To measure similarity, we employed cosine distance in the

Vector-Space model (Manning et al., 2008). Based on the calculated similarities, urban planners are able to order comments by relevance to the topic, filtering out off-topic posts. For example, there were over 400 comments with zero similarity to the topic (i.e., they did not have words in common with our topic representation). As for the sentiment expressed in the most relevant comments, the average for the first quartile shares a similar slightly negative (mean -0.13) tendency of the non-filtered comments.

**Discussion and Implications**

*Potential and Challenges in Identifying Functional Areas*

The identification of popular functional areas, popular areas for craft beer in our case, is important because these are phenomena happening in the city spontaneously, with target consumers already visiting these areas. Note the potential for the identification and exploration of other types of phenomena in the city on any thematic.

Since the originally proposed Beer Street area was identified by the analysis, this is the first evidence that social media data can be effectively used in planning as a way to support decision-making. Based on the obtained evidence, a planning decision can be made, and this decision can be evaluated. While traditionally this is often done through interviews and participatory hearings, we now could explore people's opinions shared on social media. This is a less explored point in the literature; to the best of our knowledge, we did not find any study on this topic. In this study, we show how comments of users on the news websites can be explored for that purpose. This can provide important hints on how favorable citizens are regarding a project. It is also worth mentioning that we can also analyze multiple pieces of evidence simultaneously, which could improve our understanding of a certain decision or project under evaluation.

The evidence identification step may employ existing information retrieval and data mining techniques (Han et al., 2011; Tan et al., 2006; Manning et al., 2008). However, new proposals could also be applied if, for a specific problem, an existing technique is insufficient. In fact, the discovery and exploration of functional areas using social media data were subject of investigation in previous works (Salas-Olmedo et a., 2018, Vaca et al., 2015; Wakamiya et al., 2011). In this study, we proposed an approach to extract popular areas for craft beer consumption (our evidence) using a density-based clustering algorithm. We also demonstrated how to increase the understanding of the identified areas for craft beer consumption using an official dataset about citizens' complaints and requests. This illustrates the benefit of data integration, a key point for obtaining richer knowledge of any aspect under study (Silva et al., 2019).

*Where Does Social Media Data Fit in Urban Planning?*

New data sources, such as social media data, can be used by cities in the same way as traditional datasets. The power of social media comes with their quantity and opportunities to measure unseen phenomena. For example, large amounts of social media data can speed up the evaluation process of planning as large quantities of citizens' perceptions can be measured in a short time period. Also, these new data sources provide cities with data that is not limited to bounded places or territories. Social media helps in a cost effective way to study urban phenomena and gives cities a new opportunity to monitor activities in urban areas and help decision-making.

Social media data can help the city to react and adapt to changes in the use of urban space, as shown by our case study. Quick reaction time can reduce the negative effects of certain phenomena and provides the city better opportunity to manage activities and the usage of urban space. The city of Curitiba has not yet taken advantage of emerging initiatives that have been transforming different areas inside the city. This study might be an opportunity to open the discussion in Curitiba and other cities for this new strategy. Our case study also indicates that social media data can provide information on different users and activity groups within the city that could be used, for example, in analyzing and planning tourism related services. By using new techniques, Curitiba could highlight its position as a well-planned city.

Different from conventional citizen participation methods that were incorporated into traditional planning since the 1960s, including a range of tools and tactics that require citizens to be physically present at a particular time and place (Kleinhaus et al., 2015), social media data is not dependent on the place and time. Social media also gives special groups and groups that often remain silent an opportunity to express their opinions and concerns. It is available anytime, and data typically includes different time series. In contrast to traditional data used in urban planning, social media provides opportunities for cities to be more reactive in the decision making process (Townsend, 2000). Social media data could be suitable for fast analysis supporting targeted decision making, especially in small spatial scales such as the neighborhood or street scale.

Analyzing citizens' reactions, expressed as comments, can provide clues on how favorable the population is and whether there is any misconception regarding a pre-defined project. Based on the findings, urban planners can refocus the planning if necessary or create educational campaigns to better inform the population about the project. However, manually keeping track and analyzing all social media interactions related to a project is difficult, mainly due to the sheer amount of data produced by social media users. It is important, therefore, to provide urban planners with tools to automatically assess citizens' reactions and to filter relevant comments for further consideration.

One important argument for using these new data sources is their availability and ability to use them without significant investments. However, even if the social media and new data sources can provide comparative advantage related to the cost associated with specific planning tasks, some studies, e.g., Kleinhaus et al. (2015), have found that using social media and new technologies will not reduce the workload of professionals. Our results indicate that collecting and categorising data might take significant time and expertise, and knowledge about local context and its characteristics is invaluable. Therefore, new ways of communicating and collecting data from urban dwellers can even increase the workload related to analyzing and understanding complex urban issues.

Apart from all the appeals regarding social media data, there are still some issues remaining, and they should be taken into consideration in any urban planning application. One crucial challenge is how data is covering the population or the possible bias associated. According to Blank (2016), there are substantial demographic differences between users and not users of Twitter, for instance. Our data may reflect the behavior of a fraction of consumers and, therefore, there could be biases related to the fact that the users of such applications might not represent all craft beer drinkers.

*New Data Sources Transforming Planning Practices?*

In this study, we show that combining social media data to conventional official datasets might help to fulfill the knowledge gaps the city has related to specific urban issues. In the best case, social media data could help the city to better sense and organize itself. It is important to keep in mind that other data sources, such as business licenses and accidents, could also be used as a new layer of evidence to improve the urban planning processes.

Spatial information can help the city to evaluate the characteristics chosen to be used in planning. Detailed data from social media has the potential to help the city to better organize strategies for urban planning in various territorial scales, which can make a substantial difference in the effectiveness of, for instance place branding or tourism planning (Salas-Olmedo et al., 2018; Syssner, 2010). Social media data provides a new cost effective way to measure citizens' perceptions about the city or particular activities taking place in certain neighborhoods, and so can be valuable in monitoring the effects of place branding strategies (Sevin, 2016) and citizens' perceptions on on-going planning projects. As a previous study (Silva and Graeml, 2016) shows, Curitiba is one of the most fruitful cities for craft beer in Brazil. Beer culture is, therefore, a feature the city could explore in planning initiatives. The analysis of Untappd data helps the city to determine which geographic units in the city (e.g., neighborhoods) would be suitable to develop around the craft beer activities and nightlife and rank them according to their potential.

Our data and results indicate the planning of the Beer Street could have been supported with the help of a similar approach to the one presented in this study. Social media data does not only help to confirm that Curitiba is an important city related to craft beer in Brazil but also which neighborhoods deserve more attention. As a consequence, the neighborhood where the Beer Street is planned, strategic management is essential. It can potentially help to change its current reality and bring better conditions and opportunities for its businesses and dwellers. To ensure the success of the Beer Street branding in the intended area, the city must tackle some important issues. For instance, parking and street blocking is the most popular type of complaint on cluster *H* (see Section Results). This shows that solutions to minimize these types of complaints will need to be met if the Beer Street is built on that spot.

The data acquisition step is where all contents are gathered to create the underlying database. This step is project-specific since each project may have unique purposes. Therefore, the techniques to guide the acquisition step are dependent on how the information related to the project is propagated. In any case, however, the tools involved in the process are traditional data extraction algorithms, some of them illustrated in Silva et al. (2016), with eventual manual or semi-automatic intermediate steps.

**Conclusions**

Benefits of social media data, compared to conventional official datasets, include large quantity, ability to conduct proactive problem solving, new time horizon to urban dynamics and ability to measure urban phenomena in very detailed spatial scale. The results of our case study show that social media data provides information on how certain parts of the city are used, and how this data can help enrich information used by planning authorities. Also sentiment analysis of social media responses offers chances to evaluate decision-making. In our analysis, we were able to identify the recently planned Curitiba's Beer Street from data collected from Untappd. This highlights that in this particular planning decision, the city could have used social media data as evidence to support decision-making. Furthermore, finding characteristics describing certain areas within the city can help the city organize urban activities and provide tools for urban planning.

Curitiba is well known globally for its urban planning, but how the city is taking new responsive techniques into consideration requires more research. Further research on the use of data in the planning of Curitiba, would help us to determine better the opportunities the different new data sources, such as social media data, offer to the city and how the city could continue building its reputation as a model city of urban planning.

**Funding**

This study was financed in part by the Coordenação de Aperfeiçoamento de Pessoal de Nível Superior - Brasil (CAPES) - Finance Code 001. This work is also partially supported by the project URBCOMP (Grant \#403260/2016-7 from CNPq agency) and GoodWeb (Grant \#2018/23011-1 from São Paulo Research Foundation - FAPESP). The authors would like to thank also the research agencies Fundação Araucária, and FAPEMIG.

**Disclosure**

No potential conflict of interest was reported by the authors.

**Figures**

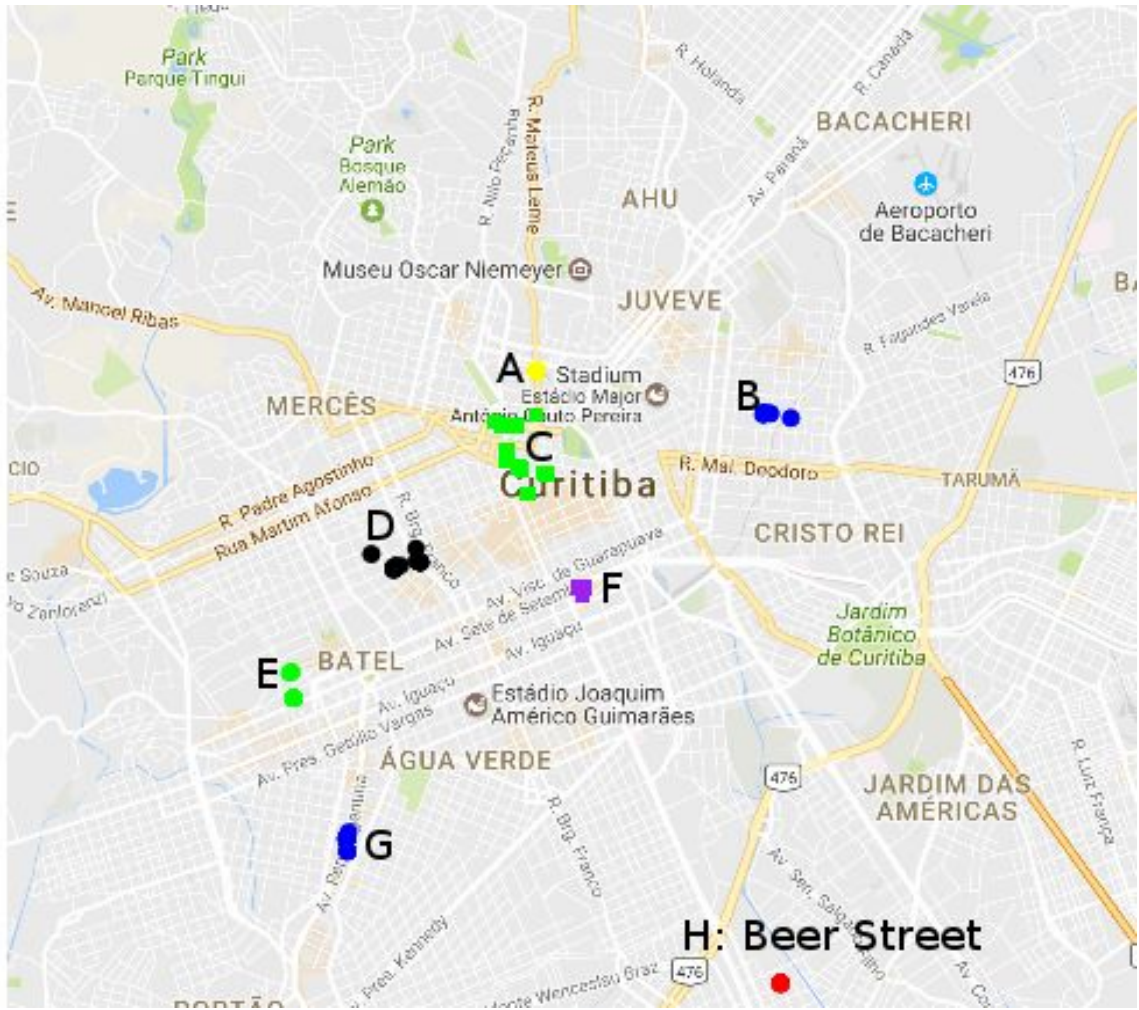

Figure 1. Clusters found in Curitiba using Untappd data. Cluster H is where the city is already planning to implement the Beer Street.

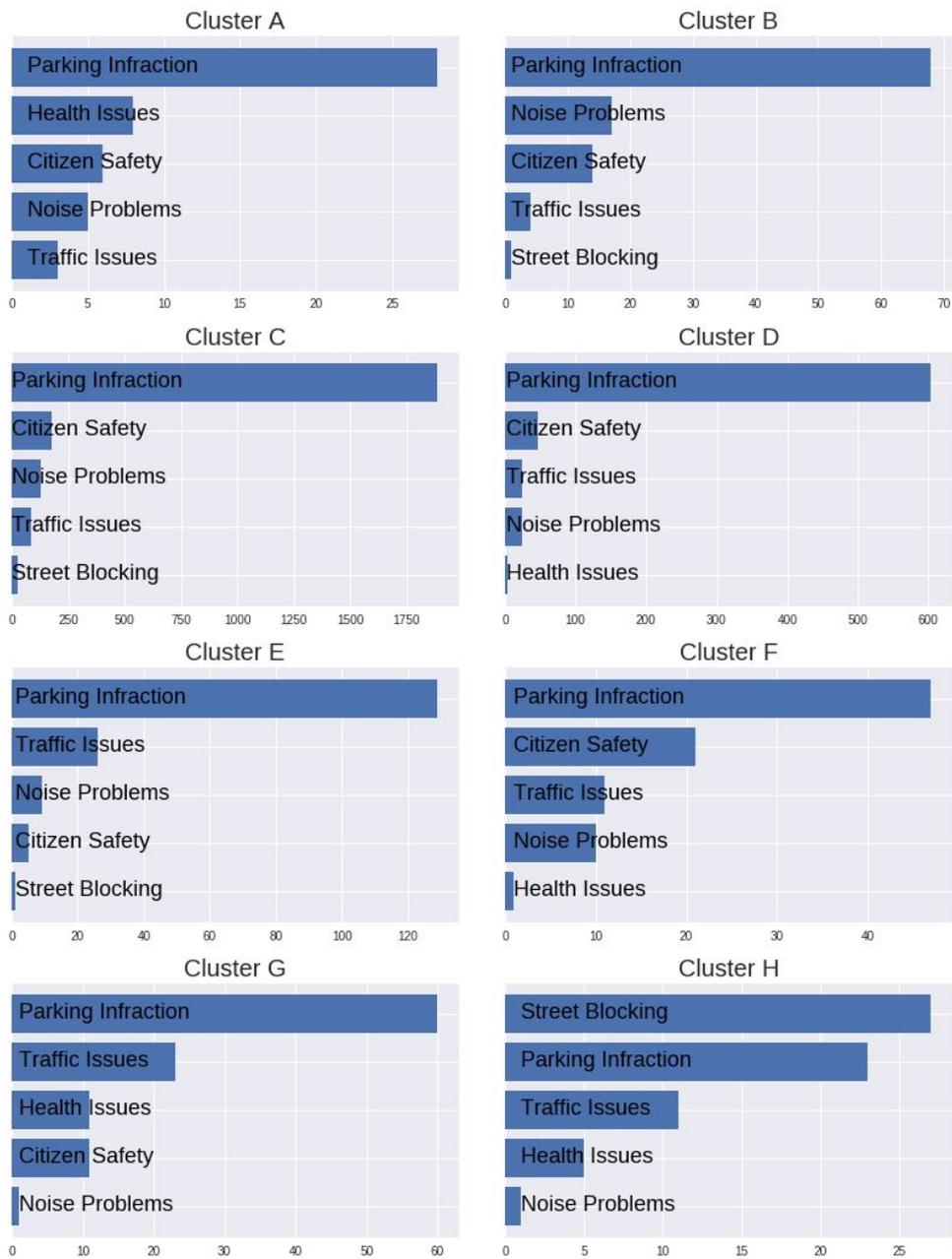

Figure 2. Distribution of the types of citizens' reports in each cluster.

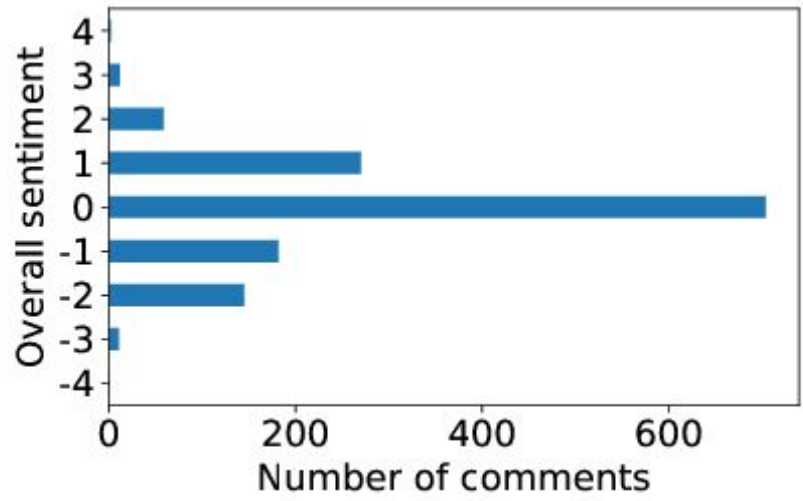

Figure 3. Distribution of overall sentiment for all studied comments.

**Tables**

Table 1: Meta-categories created for the subdivisions of reports from the dataset Reports.

| Meta-category | Examples of categories |
|---|---|
| Street Blocking | Events blocking streets; blocking street with vehicle |
| Traffic Issues | Traffic inspection; speed excess; |
| Health Issues | Leptospirosis risk; open sewage |
| Noise Problems | Loud noise at night; loud noise at daytime |
| Citizen Safety | Requesting greater security; drugged people on the street |
| Parking Infractions | Parking on sidewalk; parking at forbidden time |

Table 2. Summarization of all problems found in the studied areas and features that are not met. Assigned cell means that a certain problem is present in the cluster or the cluster does not meet a certain feature.

| Cluster | $F_1$ | $F_2$ | $F_3$ | $F_4$ | Safety | Noise | Street Block | Parking | Health |
|---|---|---|---|---|---|---|---|---|---|
| **A** | x | x |  |  | x | x |  | x |  |
| **B** |  |  | x | x | x | x | x | x |  |
| **C** |  |  | x | x | x | x | x | x |  |
| **D** |  |  | x | x | x | x |  | x | x |
| **E** |  |  | x | x | x | x |  | x | x |
| **F** | x |  |  |  | x |  |  | x | x |
| **G** | x | x |  |  | x |  |  | x | x |
| **H** |  |  |  |  |  |  | x | x | x |